\documentclass[twocolumn, prb, aps, 10pt, raggedbottom,%
superscriptaddress, nobalancelastpage]{revtex4-1}

\usepackage{graphicx}
\usepackage{amsmath}

\newcommand{\dvec}[1]{\boldsymbol{#1}}
\newcommand{\vk}{\dvec{\mathrm{k}}}
\newcommand{\vq}{\dvec{\mathrm{q}}}

\newcommand{\dmu}{\delta \mu}
\newcommand{\mubar}{\bar{\mu}}
\newcommand{\Veh}{V_{\mathrm{eh}}}
\newcommand{\Dzero}{\tilde{\Delta}}
\newcommand{\ang}[1]{\langle#1\rangle}
\newcommand{\etal}{{\textit{et al.}} }

\begin{document}

\title{Density fluctuation effects on the exciton condensate in
double layer graphene}
\author{D. S. L. Abergel}
\affiliation{Condensed Matter Theory Center, Department of Physics,
University of Maryland, College Park, MD 20742, USA}
\author{R. Sensarma}
\affiliation{Condensed Matter Theory Center, Department of Physics,
University of Maryland, College Park, MD 20742, USA}
\affiliation{Department of Theoretical Physics, Tata Institute of
Fundamental Research, Mumbai 400005, India}
\author{S. Das Sarma}
\affiliation{Condensed Matter Theory Center, Department of Physics,
University of Maryland, College Park, MD 20742, USA}

\begin{abstract}
We describe the robustness of an excitonic condensate in double layer
graphene against layer density fluctuations and the associated 
charge inhomogeneity, and discuss the implications for
observing the condensate in current experimental conditions.
We solve the mean-field equations for a finite imbalance in the Fermi
energies in each layer and utilize the results in two phenomenological
models for inhomogeneity. We find that the stability of the excitonic
condensate against density fluctuations is strongly dependent on the
size of the excitonic gap, and that transport experiments (such as
Coulomb drag) are promising methods for observing the condensate.
\end{abstract}

\maketitle

Excitonic condensates, where pairs of electrons and holes in a 
semiconductor-like system undergo a Bose Einstein
condensation, have been predicted to occur in bilayer semiconductor
heterostructures.\cite{snoke2000} They have previously been observed
experimentally under high magnetic fields in the quantum Hall
regime.\cite{eisenstein-nat432, eisenstein-sci305} Recent advances in
fabricating double layer graphene (DLG), where a layer of dielectric is
sandwiched between two graphene layers, have motivated theoretical
predictions of the emergence of an excitonic condensate at zero magnetic
field in these heterostructures\cite{zhang-prb77, min-prb78} when one
layer is electron doped and the other is hole doped.
In DLG, the thick dielectric prevents any tunneling of carriers between
the two layers, while the inter-layer (attractive electron--hole)
Coulomb interaction drives the formation of an excitonic condensate.

The estimate of the critical temperature $T_c$ for excitonic
condensation varies over a wide range, even within mean-field theory,
depending on the level of screening of the Coulomb interaction considered
in the model. At high density, unscreened Coulomb interactions produce
$T_c$ approaching room temperature,\cite{min-prb78} while static
screening leads to $T_c \lesssim 0.1\mathrm{K}$.\cite{kharitonov-prb78,
*kharitonov-semscitech25, lozovik-jetp87}
Large wave vector scattering induced by short-ranged disorder further
reduces $T_c$,\cite{bistritzer-prl101, efimkin-jetp93}
while the inclusion of dynamic screening \cite{sodemann-prb85,
lozovik-ptrsa368} and
the full band structure\cite{mink-prb84} tend to favor the pairing.
The controversial nature of the size of the excitonic gap is reinforced
by the absence of any signature of the condensate in recent Coulomb drag
experiments \cite{kim-prb83} on DLG.  Since disorder scattering strongly
suppresses $T_c$,\cite{bistritzer-prl101, efimkin-jetp93} the optimal
regime to search for excitonic condensates in DLG is the relatively
low-density regime ($k_F d<1$) where $k_F$ is the Fermi wave vector and
$d$ the inter-layer separation.  But this effectively low-density
(i.e., low $k_F$) regime in graphene is susceptible to strong density
fluctuation effects (the so-called electron-hole puddles
\cite{martin-natphys4}) arising from either extrinsic (Coulomb
impurity-induced) or intrinsic (e.g., ripple-induced) charge
inhomogeneity in the system. In the current work we explore the
interesting and important question of the effect of inhomogeneous
layer density fluctuations on the excitonic condensation, emphasizing
that this is a distinct physical mechanism from impurity disorder,
\cite{bistritzer-prl101, efimkin-jetp93}
where the latter mainly hinders the condensate formation through
momentum-conservation-breaking scattering processes.
The density fluctuations may be caused by any general scalar potential
such as the Coulomb field of charged impurities or the effect of
corrugations and ripples.
But regardless of the origin of the potential, the key distinguishing
feature of our work is that we allow for imperfect nesting of the Fermi
surfaces due to the difference in chemical potential in the two layers
which is likely to be the primary manifestation of disorder in these
devices.

\begin{figure}[tb]
	\includegraphics[]{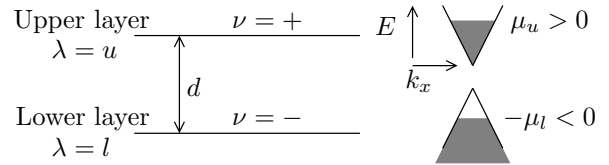}
	\caption{ Sketch of DLG doped such that the upper
	layer contains electrons and the lower layer contains holes with 
        different densities.
\label{fig:sketch}}
\end{figure}

It is well known that the charge landscape of graphene mounted on a
substrate is inhomogenous 
\cite{martin-natphys4, zhang-natphys5, deshpande-prb79} and that when
the overall density is small, ``puddles'' of electrons and holes are
formed.
Various mechanisms for the formation of this inhomogeneity have been
proposed, including the presence of charged impurities in the
environment \cite{rossi-prl101, polini-prb78} and ripples.
\cite{vazquezdeparga-prl100, gibertini-prb85}
These density fluctuations can also be viewed as inhomogeneities in the
local chemical potential.
\footnote{The width of the density fluctuations in graphene vary
\cite{xue-natmat10} from $\sim 10^{10}\mathrm{cm}^{-2}$ (best samples on
hBN) to $\sim 10^{11}\mathrm{cm}^{-2}$ (on SiO$_2$).}  
The presence of the top and bottom gates, as well as the dielectric
between the graphene layers, implies that such chemical potential
fluctuations are ubiquitous in DLG samples. 
We study the effects of these fluctuations on the
excitonic condensate within a local density approximation. 
Keeping in mind the wide variation of the excitonic gap in the clean
case based on the theoretical approximation scheme, we look at the 
two extreme cases: 
(i) unscreened Coulomb interaction and
(ii) Coulomb interaction with static screening, with a focus on common
features obtained by scaling various energy scales by
$\tilde{\Delta}_0$, which is the excitonic gap for the clean sample in
either approximation. 

Our main results are the following: (a) In spatially homogeneous DLG,
with different electron and hole Fermi energies, the pairing profile
$\Delta_k$ depends only on the average Fermi energy $\mubar$, and is
independent of the difference $\dmu$ until
$|\delta\mu|=2\tilde{\Delta}_0$. 
Beyond this point, the pairing collapses for both the screened and the
unscreened interaction. 
(b) For imbalanced Fermi surfaces, we find no evidence of a
Fulde-Ferrell-Larkin-Ovchinnikov (FFLO) type state \cite{fulde-pr135}
with pairing at finite center-of-mass momentum $|\dvec{\mathrm{Q}}| =
k_{Fu}-k_{Fl}$ in either the unscreened or the screened case.
(c) We provide two different estimates of the effect of charge
inhomogeneity on the excitonic condensate:
(i) the average pairing gap, which is relevant for thermodynamic
measurements, and (ii) the fraction of the sample area supporting the
condensate which is relevant
for transport measurements. We discuss these quantities as a function of
the average chemical potential and the width of density fluctuations
for both screening models. 
Our results show that screening makes the excitonic condensates more
vulnerable to charge fluctuations by virtue of the smaller excitonic gap 
and that transport measurements are more robust to these fluctuations
than bulk thermodynamic measurements.

The large separation of the two graphene layers implies that the local
chemical potential (and hence charge) fluctuations are independent of
each other. For convenience, we will assume the top layer to be electron
doped with local chemical potential $\mu_u$ 
and the lower layer to be hole doped with local chemical potential
$-\mu_l$. These quantities may fluctuate spatially. 
The independent fluctuations of the two chemical potentials imply that
the perfect nesting of Fermi surfaces is lost in these inhomogeneous
materials even when the global average $\langle\mu\rangle$ is the same 
in each layer.
Before going into the details of averaging over distributions of
chemical potentials through a local density approximation, we 
examine the effect of unequal (but spatially homogeneous) 
chemical potentials in
the two layers on the excitonic properties.

The Hamiltonian for a single spin/valley species in monolayer graphene
in layer $\lambda \in \{u,l\}$ is given by $H_{\lambda\nu} =
\sum_{\vk} \left( \varepsilon_{\vk\nu}-\nu \mu_{\lambda} \right)
c^{\dagger}_{\lambda\vk\nu} c_{\lambda\vk\nu}$ where $\nu \in \{+,-\}$
denotes the band and
$\varepsilon_{\vk\nu} = \nu v k$ is the single particle energy. We
will set $\hbar=1$ for the rest of this paper.  For excitonic
condensates in DLG, we are interested in the case
where the average electron (hole) doping in the upper (lower) layer is
large (i.e., $ |\mu_\lambda| \gg \Delta_0$). Since exciton condensation
is dominated by processes around the Fermi surfaces, we can exclude
the filled (empty) valence (conduction) band in the upper (lower)
layer from our model. This is illustrated in Fig.~\ref{fig:sketch}.
The full Hamiltonian is then $H = H_{u+} + H_{l-} + \Veh$ where $\Veh$
is the inter-layer interaction between the electrons and
holes. We neglect the intra-layer interaction since it amounts only to
a small renormalization of the Fermi velocity in the layer.
\cite{dassarma-prb75, dassarma-arXiv1203} We write
the second-quantized operators for electrons in the upper layer as
$a_{\vk}= c_{u\vk+}$ and for holes in the lower layer as $b_{\vk} =
c^{\dagger}_{l\vk-}$. This particle-hole transformation allows us to
make a direct connection with the BCS theory of superconductivity,
albeit with a complicated and non-separable potential kernel. In this
notation, the BCS pairing ansatz with zero center-of-mass momentum for
the interaction gives
\begin{equation}
	\Veh = \sum_{\vk\vq} V(\vq) f(\vk+\vq,\vk)
	a^{\dagger}_{-\vk-\vq} b^{\dagger}_{\vk+\vq} b_{\vk} a_{-\vk}
\end{equation}
where $V(\vq)$ is the interaction potential and $f(\vk,\vk') = [1 +
\cos(\theta_k-\theta_{k'})]/2$ is the chirality factor due to 
the graphene band wave functions. Using standard
diagonalization techniques with a mean field in the excitonic channel,
$\Delta_{\vk} = \ang{ b_{\vk} a_{-\vk}}$, we obtain two quasiparticle
branches with spectrum $E_{\alpha,\beta} = \dmu/2
\pm \sqrt{ \left( v k - \mubar \right)^2 + \Delta_{\vk}^2}$, where
$\mubar=(\mu_u+\mu_l)/2$ and $\dmu=\mu_u-\mu_l$. The
self-consistent gap equation is then
given by
\begin{equation}
	\Delta_{\vk} = \sum_{\vk'} V(\vk'-\vk)
	\frac{ \Delta_{\vk'} f(\vk,\vk') 
	\left[ n_{\beta}(k') - n_{\alpha}(k') \right]}{ \sqrt{
		(vk' - \mubar)^2 + \Delta_{\vk'}^2}}
	\label{eq:gap}
\end{equation}
where the occupation numbers are $n_{\alpha(\beta)}(k)=
\Theta(-E_{\alpha(\beta)}(k))$ at zero temperature.  Equation
\eqref{eq:gap} shows that  $\mubar$ and $\dmu$ are the natural variables
for analyzing the system.  We
will first look at the system with $\dmu=0$, where the two Fermi
surfaces are perfectly nested.

\begin{figure}[tb]
	\includegraphics[]{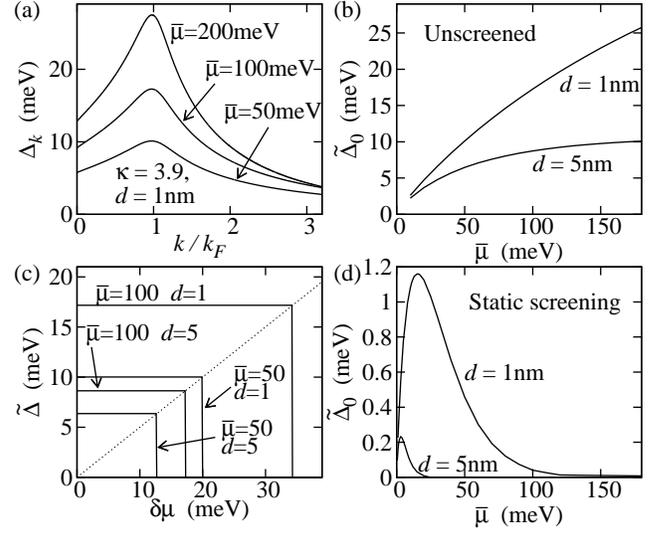}
	\caption{(a) The pairing amplitude $\Delta_{\vk}$ for unscreened
	interaction with $\dmu=0$ and different $\mubar$. 
	(b) The largest pairing amplitude $\tilde{\Delta}_0$ for the
	unscreened interaction at $\dmu=0$.
	(c) The evolution of the largest pairing amplitude $\Dzero$ with
	$\dmu$ for the unscreened interaction. The dotted line is $\dmu / 2$
	which serves as a guide to the eye illustrating the collapse of the
	excitonic pairing when $\dmu = 2\Dzero_0$.
	(d) $\tilde{\Delta}_0$ for the statically screened interaction at
	$\dmu=0$. 
	\label{fig:gaps}}
\end{figure}

Within a model of unscreened Coulomb interaction,\cite{zhang-prb77,
  min-prb78} which gives the most generous estimate for the
excitionic gap, the inter-layer potential is $V(\vq) \equiv V_b(q)
e^{-qd} = 2\pi e^2 e^{-qd} / (\kappa q)$ where $V_b(q)$ is the bare
two-dimensional Coulomb interaction and $\kappa$ is the dielectric
constant of the
environment. \footnote{Throughout this paper, we will use $\kappa=3.9$
corresponding to the dielectric constant for hBN or SiO$_2$.} The
momentum profile of $\Delta_k$ for $\dmu = 0$ is shown in
Fig.~\ref{fig:gaps}(a) for a small inter-layer separation of $1$nm and
three different values of $\mubar$.  $\Delta_k$ has a non-monotonic
momentum dependence with a peak at $k=k_F$ whose height increases with
the size of the Fermi surface. This peak height, $\tilde{\Delta}_0$,
is shown in Fig.~\ref{fig:gaps}(b) as a function of $\mubar$ for two
different values of the inter-layer separation,
$d=1,5\mathrm{nm}$. The profiles have a convex shape which flattens
out at large values of $\mubar$. For $\dmu=0$, the quasiparticle
spectrum $E_\alpha=-E_\beta$ and the positive energy branch has a
minimum $\tilde{\Delta}_0$ at $k=k_F$. Thus $\tilde{\Delta}_0$ can be
identified as the true single particle spectral gap in the system.

In the opposite limit, we study the model of Coulomb interactions with
static screening, which gives the lowest estimate of excitonic
gaps at high density. In this case, the inter-layer potential is given
by \cite{lozovik-jetp87}
\begin{equation*}
	\Veh(\vq) = \frac{ V_b(q) e^{-qd}}
	{1 - V_b(q)\left[ \Pi_u + \Pi_l \right] + V_b(q)^2 \Pi_u \Pi_l
	\left( 1 - e^{-2qd} \right) }
\end{equation*}
where $\Pi_\lambda(q)$ is the polarization function in a single layer
\cite{hwang-prb75} which may be different in the two layers for
mismatched Fermi surfaces. The momentum dependence of $\Delta_k$ takes
a similar form to the unscreened case, with a peak at $k=k_F$, but, as
shown in Fig.~\ref{fig:gaps}(d), the size of the peak
$\tilde{\Delta}_0$ is several orders of magnitude smaller. Unlike the
unscreened case, $\tilde{\Delta}_0$ shows a non-monotonic
dependence on $\mubar$ with a peak occuring at a relatively small
value of $\mubar$. The increase in $\tilde{\Delta}_0$ at small
$\mubar$ is due to the increasing density of states at the Fermi energy,
but increasing the size of Fermi surface also makes the screening of the
Coulomb interaction more efficient since the Thomas Fermi wave vector
$q_{TF} \sim k_F$ in linearly dispersing
graphene. At large densities, the increased screening overwhelms the
increasing density of states and the excitonic gap decreases rapidly.
\cite{kharitonov-prb78, lozovik-jetp87}

We now focus our attention on the effects of Fermi surface imbalance
on the excitonic properties, using a pairing ansatz
with zero center of mass momentum.
We see that changing the sign of $\dmu$ is equivalent to changing
$E_\alpha \rightarrow -E_\beta$ so a discussion of $\dmu >0$ will
suffice.  As $\dmu$ increases, we find that the profile of $\Delta_k$
remains unchanged until $\dmu=2\tilde{\Delta}_0$. For
$\dmu>2\tilde{\Delta}_0$, the pairing vanishes for all $k$ leading to
the normal state.  This behavior is common to both the unscreened and
screened interaction model and is equivalent to the
Clogston-Chandrasekhar \cite{clogston-prl9} limit for loss of pairing
in a superconductor in a Zeeman field.  This is visible explicitly in
Fig.~\ref{fig:gaps}(c), where the peak pairing amplitude
$\tilde{\Delta}$ is plotted as a function of $\dmu$ for the unscreened
interaction for $\mubar=50$ and $100$meV and $d=1$ and $5$nm.
This behavior can be understood from Eq.~\eqref{eq:gap}
where $\dmu$ appears only through the Fermi
functions $n_{\alpha(\beta)}$. 
The profile for $\Delta_k$ is unchanged from the $\dmu=0$ case so the
occupation probabilities are unchanged if $|\dmu|/2<\tilde{\Delta}_0$.
Thus the profile for $\dmu=0$ satisfies the gap equation for $\dmu <
2\tilde{\Delta}_0$. Beyond this value,
the occupation probabilities are changed around the gap edge, where
the gap edge singularities in the density of states ensure that the
pairing is completely lost.

We have also looked at the possibility of electron-hole pairing with
finite center of mass in the case of imbalanced Fermi surfaces,
specifically for pairing with center of mass momentum
$|Q|=|k_{Fu}-k_{Fl}|$, which has a rich history in the theory of
superconductivity under the name ``FFLO pairing''.\cite{fulde-pr135}
We have not found any evidence of FFLO type excitonic pairing in either
the screened or the unscreened model. This is not surprising, since,
even with the model of a constant local interaction, FFLO states are
known to stabilize in only a narrow parameter regime in higher
than one dimension.\cite{sheehy-annphys322,*conduit-pra77}
The momentum dependence of the Coulomb potential would further
destabilize this fragile state.
We safely predict the non-existence of any FFLO-type interlayer
excitonic superfluidity in density-imbalanced bilayer systems.

\begin{figure}[tb]
	\includegraphics[]{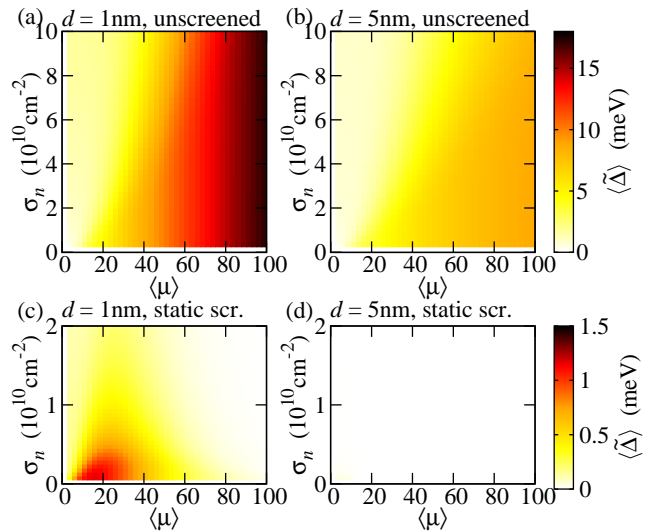}
	\caption{Areally averaged excitonic gap, given by
	Eq.~\eqref{eq:avDel}. The layer separation and screening model are
	labeled.
	\label{fig:average}}
\end{figure}

Having looked at the effects of Fermi surface imbalance on homogeneous
states, we now connect these findings to excitonic condensates in
actual inhomogeneous DLG samples. 
In the first instance, one could imagine a thermodynamic experimental
probe which couples to an area of the sample which is larger than the
size of the average fluctuation (e.g., measurement of a gap derived from
specific heat). 
In this case, the measured excitonic gap might be best modeled as an
average of the gap over the spatial area sampled. 
Since the correlation length of the density fluctuations is of the order
of the system size, we can use a local density approximation to write
\begin{equation}
	\ang{\tilde{\Delta}} = \iint d\mu_l \, d\mu_u  \,
	P( \mu_u )
	P( \mu_l )
	\tilde{\Delta}(\mu_u,\mu_l)
	\label{eq:avDel}
\end{equation}
where $P(\mu_\lambda)$ is the probability distribution for the
chemical potential in layer $\lambda$. 
Similar averaging procedures have been employed with great success.
\cite{abergel-prb83}
It is well known that for monolayer graphene, the distribution of
density fluctuations due to charge impurities is
Gaussian, \cite{rossi-prl101, xue-natmat10} and hence
\begin{equation}
	P(\mu) = \frac{2|\mu|}{\sqrt{2\pi^3}v^2\sigma_n} 
	\exp\left[
	-\frac{(\mu|\mu|-\ang{\mu}|\ang{\mu}|)^2}{2\sigma_n^2v^4\pi^2}\right]
\end{equation}
where $\sigma_n$ is the width of density fluctuations determined by
impurity concentrations, $\ang{\mu}$ is the global chemical potential
(which we assume to be the same for both layers and is set by external
gating). 
It is known from theoretical work \cite{rossi-prl101} that in the case
of charged impurities, $\sigma_n$ and the two-dimensional impurity
concentration are of the same order of magnitude. 
If the density fluctuations are caused by ripples, there is no reason to
believe that these should be the same in the two layers; on the other
hand, the field due to charged impurities will be effectively screened
from the more distant layer by the closer one.  Hence we assume that the
inhomogeneities in the two layers are uncorrelated.
However, inter-layer correlations can be included by substituting a more
sophisticated function instead of the product $P(\mu_u) P(\mu_l)$.
Thus we have reduced the effects of inhomogeneity to one
phenomenological parameter $\sigma_n$ which takes the same value in
both layers. 
Evaluating the integrals in Eq.~\eqref{eq:avDel} gives the color plots
shown in Fig.~\ref{fig:average}. In the unscreened case for small
inter-layer distance, the gap is large enough that the presence of
density fluctuations does not substantially alter the gap. However,
for static screening, the fluctuations are large enough to
completely kill the exciton condensate for moderate inter-layer
separation, and even for minimal separation at low density a vanishing
amount of inhomogeneity is required if the gap is to persist.

\begin{figure}[tb]
	\includegraphics[]{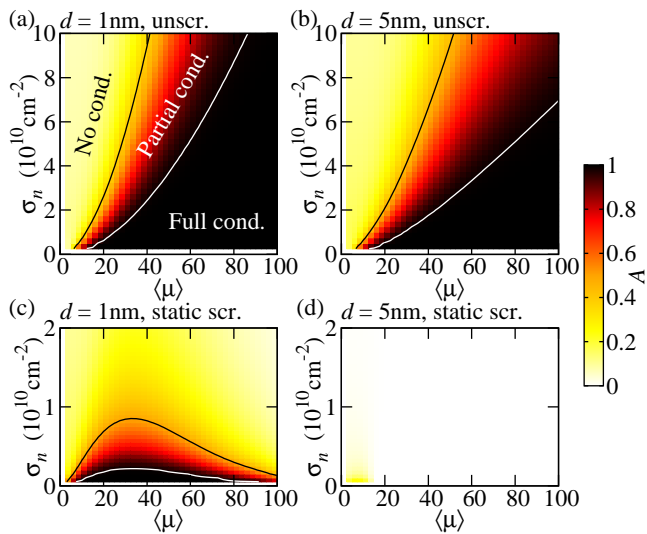}
	\caption{Fraction of the area supporting the excitonic
	condensate in inhomogeneous DLG. The black contour
	denotes $A = 0.5$ and the white contour is $A=1$.
	\label{fig:percolation}}
\end{figure}

A more definitive signature of excitonic condensates is the
superfluid or perfectly inductive response of the system in drag
transport measurements. 
The condensate and non-condensate regions form a network, and
percolation theory can be applied to determine the transport
properties.\cite{rossi-prl107} The strongly inhomogeneous samples most
probably show glassy behavior, but the nature of this glassy state
requires a more sophisticated treatment of disorder.
Although a full analysis of percolating clusters is beyond the scope of
this paper, the fraction of area supporting the condensates can be
related to percolation thresholds.
The fractional area of the sample which supports the condensate is
given by
\begin{equation}
	A = \iint d\mu_u\, d\mu_l  P(\mu_u) P(\mu_l)
	\Theta(\mu_u\mu_l) \Theta(2\Dzero_0-|\dmu|)
\end{equation}
The first step function represents the fact that in the inhomogeneous
case it is possible for $\mu_u$ and $\mu_l$ to be in the same band. If
this is the case then the condensate is not allowed.
Figure~\ref{fig:percolation} shows $A$ for both the unscreened and
statically screened interactions, and for the two
values of the inter-layer separation shown previously.
The black contour denotes $A=0.5$, where half the area supports the
condensate, and the white contour shows $A=1$ meaning that the whole
sample is excitonic. We see that in the unscreened case,
the condensate should be stable at experimentally accessible values of
density fluctuations, but in the statically screened case the
fluctuations have to be smaller even than those currently available in
hexagonal boron nitride (hBN) devices.\cite{dean-natnano5, xue-natmat10}
By comparing Figs.~\ref{fig:percolation} and \ref{fig:average}, we see
that in samples with density fluctuations, the formation of the
excitonic condensate is easier to detect in transport-style experiments
than in probes that average over the bulk of the sample.
\footnote{Such experiments could benefit enormously if the inhomogeneity
in the individual graphene layers could be screened out in a controlled
manner by using screening layers as was done in a clever recent
experiment by Ponomarenko \etal.\cite{ponomarenko-natphys7}}

We also note that other authors \cite{sodemann-prb85, lozovik-ptrsa368} 
have considered the effect of dynamic screening on the exciton
condensate without disorder. 
In the high-density limit (where static screening gives a tiny gap due
to the large size of the polarization function), dynamic screening 
gives a gap which is closer to the unscreened interaction than the static
screening.
Using the data presented in Ref.\,\onlinecite{sodemann-prb85}, we
estimate that for physically realistic parameters, the dynamic screening
theory predicts that the peak gap in the high-density regime
will be $\sim 0.1\mathrm{meV}$,
indicating that the numerical effect of disorder is roughly similar to
that of the static screening in the low density regime. Since our
analysis shows that the nature of the screening has no impact on the
effects of inhomogeneity, Figs. \ref{fig:average} and
\ref{fig:percolation} show that a gap of this size is marginally
observable.
Therefore, just as the low density statically screened interaction may
be observed in the cleanest systems, so should a gap generated by
dynamic screening in the high-density regime.

In summary, we have examined the impact of chemical potential
fluctuations and charge inhomogeneity on the formation of an excitonic
condensate in DLG. The origin of these charge fluctuations may be from
ripples in the graphene surface, charged impurities in the environment
of the graphene, or any other general spatially-varying scalar field.
While the absolute value of the gap depends sensitively on the model of
interaction used, the effects of Fermi surface imbalance are identical
in the two cases. 
However, since the inhomogeneity can be controlled only on an
absolute scale, we find that the size of the gap is the crucial factor
which determines the robustness of the condensate against this form of
disorder. If the gap is of the order of a few meV, then the level of
fluctuations found with hBN substrates should be sufficient to observe
the condensate in Coulomb drag experiments. Finally, if static screening
is important in the formation of excitonic condensates, there is an
optimal Fermi energy (which is $\sim 20\mathrm{meV}$ for
$d=1\mathrm{nm}$), where it would be easiest to see the condensate. 

\begin{acknowledgments}
This work is supported by NRI-SWAN and ONR-MURI.  
D.S.L.A. and R.S. acknowledge the KITP, supported in part by the
National Science Foundation under Grant No. PHY11-25915, where part of
the work was carried out.
\end{acknowledgments}

\bibliography{bibtex-sorted}

\begin{thebibliography}{36}%
\makeatletter
\providecommand \@ifxundefined [1]{%
 \@ifx{#1\undefined}
}%
\providecommand \@ifnum [1]{%
 \ifnum #1\expandafter \@firstoftwo
 \else \expandafter \@secondoftwo
 \fi
}%
\providecommand \@ifx [1]{%
 \ifx #1\expandafter \@firstoftwo
 \else \expandafter \@secondoftwo
 \fi
}%
\providecommand \natexlab [1]{#1}%
\providecommand \enquote  [1]{``#1''}%
\providecommand \bibnamefont  [1]{#1}%
\providecommand \bibfnamefont [1]{#1}%
\providecommand \citenamefont [1]{#1}%
\providecommand \href@noop [0]{\@secondoftwo}%
\providecommand \href [0]{\begingroup \@sanitize@url \@href}%
\providecommand \@href[1]{\@@startlink{#1}\@@href}%
\providecommand \@@href[1]{\endgroup#1\@@endlink}%
\providecommand \@sanitize@url [0]{\catcode `\\12\catcode `\$12\catcode
  `\&12\catcode `\#12\catcode `\^12\catcode `\_12\catcode `\%12\relax}%
\providecommand \@@startlink[1]{}%
\providecommand \@@endlink[0]{}%
\providecommand \url  [0]{\begingroup\@sanitize@url \@url }%
\providecommand \@url [1]{\endgroup\@href {#1}{\urlprefix }}%
\providecommand \urlprefix  [0]{URL }%
\providecommand \Eprint [0]{\href }%
\providecommand \doibase [0]{http://dx.doi.org/}%
\providecommand \selectlanguage [0]{\@gobble}%
\providecommand \bibinfo  [0]{\@secondoftwo}%
\providecommand \bibfield  [0]{\@secondoftwo}%
\providecommand \translation [1]{[#1]}%
\providecommand \BibitemOpen [0]{}%
\providecommand \bibitemStop [0]{}%
\providecommand \bibitemNoStop [0]{.\EOS\space}%
\providecommand \EOS [0]{\spacefactor3000\relax}%
\providecommand \BibitemShut  [1]{\csname bibitem#1\endcsname}%
\let\auto@bib@innerbib\@empty
\bibitem [{\citenamefont {Moskalenko}\ and\ \citenamefont
  {Snoke}(2000)}]{snoke2000}%
  \BibitemOpen
  \bibfield  {author} {\bibinfo {author} {\bibfnamefont {S.~A.}\ \bibnamefont
  {Moskalenko}}\ and\ \bibinfo {author} {\bibfnamefont {D.~W.}\ \bibnamefont
  {Snoke}},\ }\href@noop {} {\emph {\bibinfo {title} {Bose-Einstein
  Condensation of Excitons and Biexcitons}}}\ (\bibinfo  {publisher} {Cambridge
  University Press, Cambridge},\ \bibinfo {year} {2000})\BibitemShut {NoStop}%
\bibitem [{\citenamefont {Eisenstein}\ and\ \citenamefont
  {MacDonald}(2004)}]{eisenstein-nat432}%
  \BibitemOpen
  \bibfield  {author} {\bibinfo {author} {\bibfnamefont {J.~P.}\ \bibnamefont
  {Eisenstein}}\ and\ \bibinfo {author} {\bibfnamefont {A.~H.}\ \bibnamefont
  {MacDonald}},\ }\href@noop {} {\bibfield  {journal} {\bibinfo  {journal}
  {Nature (London)}\ }\textbf {\bibinfo {volume} {432}},\ \bibinfo {pages}
  {691} (\bibinfo {year} {2004})}\BibitemShut {NoStop}%
\bibitem [{\citenamefont {Eisenstein}(2004)}]{eisenstein-sci305}%
  \BibitemOpen
  \bibfield  {author} {\bibinfo {author} {\bibfnamefont {J.~P.}\ \bibnamefont
  {Eisenstein}},\ }\href@noop {} {\bibfield  {journal} {\bibinfo  {journal}
  {Science}\ }\textbf {\bibinfo {volume} {305}},\ \bibinfo {pages} {950}
  (\bibinfo {year} {2004})}\BibitemShut {NoStop}%
\bibitem [{\citenamefont {Zhang}\ and\ \citenamefont
  {Joglekar}(2008)}]{zhang-prb77}%
  \BibitemOpen
  \bibfield  {author} {\bibinfo {author} {\bibfnamefont {C.-H.}\ \bibnamefont
  {Zhang}}\ and\ \bibinfo {author} {\bibfnamefont {Y.~N.}\ \bibnamefont
  {Joglekar}},\ }\href {\doibase 10.1103/PhysRevB.77.233405} {\bibfield
  {journal} {\bibinfo  {journal} {Phys. Rev. B}\ }\textbf {\bibinfo {volume}
  {77}},\ \bibinfo {pages} {233405} (\bibinfo {year} {2008})}\BibitemShut
  {NoStop}%
\bibitem [{\citenamefont {Min}\ \emph {et~al.}(2008)\citenamefont {Min},
  \citenamefont {Bistritzer}, \citenamefont {Su},\ and\ \citenamefont
  {MacDonald}}]{min-prb78}%
  \BibitemOpen
  \bibfield  {author} {\bibinfo {author} {\bibfnamefont {H.}~\bibnamefont
  {Min}}, \bibinfo {author} {\bibfnamefont {R.}~\bibnamefont {Bistritzer}},
  \bibinfo {author} {\bibfnamefont {J.-J.}\ \bibnamefont {Su}}, \ and\ \bibinfo
  {author} {\bibfnamefont {A.~H.}\ \bibnamefont {MacDonald}},\ }\href {\doibase
  10.1103/PhysRevB.78.121401} {\bibfield  {journal} {\bibinfo  {journal} {Phys.
  Rev. B}\ }\textbf {\bibinfo {volume} {78}},\ \bibinfo {pages} {121401}
  (\bibinfo {year} {2008})}\BibitemShut {NoStop}%
\bibitem [{\citenamefont {Kharitonov}\ and\ \citenamefont
  {Efetov}(2008)}]{kharitonov-prb78}%
  \BibitemOpen
  \bibfield  {author} {\bibinfo {author} {\bibfnamefont {M.~Y.}\ \bibnamefont
  {Kharitonov}}\ and\ \bibinfo {author} {\bibfnamefont {K.~B.}\ \bibnamefont
  {Efetov}},\ }\href {\doibase 10.1103/PhysRevB.78.241401} {\bibfield
  {journal} {\bibinfo  {journal} {Phys. Rev. B}\ }\textbf {\bibinfo {volume}
  {78}},\ \bibinfo {pages} {241401} (\bibinfo {year} {2008})}\BibitemShut
  {NoStop}%
\bibitem [{\citenamefont {Kharitonov}\ and\ \citenamefont
  {Efetov}(2010)}]{kharitonov-semscitech25}%
  \BibitemOpen
  \bibfield  {author} {\bibinfo {author} {\bibfnamefont {M.~Y.}\ \bibnamefont
  {Kharitonov}}\ and\ \bibinfo {author} {\bibfnamefont {K.~B.}\ \bibnamefont
  {Efetov}},\ }\href {http://stacks.iop.org/0268-1242/25/i=3/a=034004}
  {\bibfield  {journal} {\bibinfo  {journal} {Semicond. Sci. Tech.}\ }\textbf
  {\bibinfo {volume} {25}},\ \bibinfo {pages} {034004} (\bibinfo {year}
  {2010})}\BibitemShut {NoStop}%
\bibitem [{\citenamefont {Lozovik}\ and\ \citenamefont
  {Sokolik}(2008)}]{lozovik-jetp87}%
  \BibitemOpen
  \bibfield  {author} {\bibinfo {author} {\bibfnamefont {Y.~E.}\ \bibnamefont
  {Lozovik}}\ and\ \bibinfo {author} {\bibfnamefont {A.~A.}\ \bibnamefont
  {Sokolik}},\ }\href@noop {} {\bibfield  {journal} {\bibinfo  {journal} {JETP
  Lett.}\ }\textbf {\bibinfo {volume} {87}},\ \bibinfo {pages} {61} (\bibinfo
  {year} {2008})}\BibitemShut {NoStop}%
\bibitem [{\citenamefont {Bistritzer}\ and\ \citenamefont
  {MacDonald}(2008)}]{bistritzer-prl101}%
  \BibitemOpen
  \bibfield  {author} {\bibinfo {author} {\bibfnamefont {R.}~\bibnamefont
  {Bistritzer}}\ and\ \bibinfo {author} {\bibfnamefont {A.~H.}\ \bibnamefont
  {MacDonald}},\ }\href {\doibase 10.1103/PhysRevLett.101.256406} {\bibfield
  {journal} {\bibinfo  {journal} {Phys. Rev. Lett.}\ }\textbf {\bibinfo
  {volume} {101}},\ \bibinfo {pages} {256406} (\bibinfo {year}
  {2008})}\BibitemShut {NoStop}%
\bibitem [{\citenamefont {Efimkin}\ \emph {et~al.}(2011)\citenamefont
  {Efimkin}, \citenamefont {Kalbachinskii},\ and\ \citenamefont
  {Lozovik}}]{efimkin-jetp93}%
  \BibitemOpen
  \bibfield  {author} {\bibinfo {author} {\bibfnamefont {D.~K.}\ \bibnamefont
  {Efimkin}}, \bibinfo {author} {\bibfnamefont {V.~A.}\ \bibnamefont
  {Kalbachinskii}}, \ and\ \bibinfo {author} {\bibfnamefont {Y.~E.}\
  \bibnamefont {Lozovik}},\ }\href@noop {} {\bibfield  {journal} {\bibinfo
  {journal} {JETP Lett.}\ }\textbf {\bibinfo {volume} {93}},\ \bibinfo {pages}
  {238} (\bibinfo {year} {2011})}\BibitemShut {NoStop}%
\bibitem [{\citenamefont {Sodemann}\ \emph {et~al.}(2012)\citenamefont
  {Sodemann}, \citenamefont {Pesin},\ and\ \citenamefont
  {MacDonald}}]{sodemann-prb85}%
  \BibitemOpen
  \bibfield  {author} {\bibinfo {author} {\bibfnamefont {I.}~\bibnamefont
  {Sodemann}}, \bibinfo {author} {\bibfnamefont {D.~A.}\ \bibnamefont {Pesin}},
  \ and\ \bibinfo {author} {\bibfnamefont {A.~H.}\ \bibnamefont {MacDonald}},\
  }\href {\doibase 10.1103/PhysRevB.85.195136} {\bibfield  {journal} {\bibinfo
  {journal} {Phys. Rev. B}\ }\textbf {\bibinfo {volume} {85}},\ \bibinfo
  {pages} {195136} (\bibinfo {year} {2012})}\BibitemShut {NoStop}%
\bibitem [{\citenamefont {Lozovik}\ \emph {et~al.}(2010)\citenamefont
  {Lozovik}, \citenamefont {Ogarkov},\ and\ \citenamefont
  {Sokolik}}]{lozovik-ptrsa368}%
  \BibitemOpen
  \bibfield  {author} {\bibinfo {author} {\bibfnamefont {Y.~E.}\ \bibnamefont
  {Lozovik}}, \bibinfo {author} {\bibfnamefont {S.~L.}\ \bibnamefont
  {Ogarkov}}, \ and\ \bibinfo {author} {\bibfnamefont {A.~A.}\ \bibnamefont
  {Sokolik}},\ }\href {\doibase 10.1098/rsta.2010.0224} {\bibfield  {journal}
  {\bibinfo  {journal} {Phil. Trans. R. Soc. A}\ }\textbf {\bibinfo {volume}
  {368}},\ \bibinfo {pages} {5417} (\bibinfo {year} {2010})}\BibitemShut
  {NoStop}%
\bibitem [{\citenamefont {Mink}\ \emph {et~al.}(2011)\citenamefont {Mink},
  \citenamefont {Stoof}, \citenamefont {Duine},\ and\ \citenamefont
  {MacDonald}}]{mink-prb84}%
  \BibitemOpen
  \bibfield  {author} {\bibinfo {author} {\bibfnamefont {M.~P.}\ \bibnamefont
  {Mink}}, \bibinfo {author} {\bibfnamefont {H.~T.~C.}\ \bibnamefont {Stoof}},
  \bibinfo {author} {\bibfnamefont {R.~A.}\ \bibnamefont {Duine}}, \ and\
  \bibinfo {author} {\bibfnamefont {A.~H.}\ \bibnamefont {MacDonald}},\ }\href
  {\doibase 10.1103/PhysRevB.84.155409} {\bibfield  {journal} {\bibinfo
  {journal} {Phys. Rev. B}\ }\textbf {\bibinfo {volume} {84}},\ \bibinfo
  {pages} {155409} (\bibinfo {year} {2011})}\BibitemShut {NoStop}%
\bibitem [{\citenamefont {Kim}\ \emph {et~al.}(2011)\citenamefont {Kim},
  \citenamefont {Jo}, \citenamefont {Nah}, \citenamefont {Yao}, \citenamefont
  {Banerjee},\ and\ \citenamefont {Tutuc}}]{kim-prb83}%
  \BibitemOpen
  \bibfield  {author} {\bibinfo {author} {\bibfnamefont {S.}~\bibnamefont
  {Kim}}, \bibinfo {author} {\bibfnamefont {I.}~\bibnamefont {Jo}}, \bibinfo
  {author} {\bibfnamefont {J.}~\bibnamefont {Nah}}, \bibinfo {author}
  {\bibfnamefont {Z.}~\bibnamefont {Yao}}, \bibinfo {author} {\bibfnamefont
  {S.~K.}\ \bibnamefont {Banerjee}}, \ and\ \bibinfo {author} {\bibfnamefont
  {E.}~\bibnamefont {Tutuc}},\ }\href {\doibase 10.1103/PhysRevB.83.161401}
  {\bibfield  {journal} {\bibinfo  {journal} {Phys. Rev. B}\ }\textbf {\bibinfo
  {volume} {83}},\ \bibinfo {pages} {161401} (\bibinfo {year}
  {2011})}\BibitemShut {NoStop}%
\bibitem [{\citenamefont {Martin}\ \emph {et~al.}(2008)\citenamefont {Martin},
  \citenamefont {Akerman}, \citenamefont {Ulbricht}, \citenamefont {Lohmann},
  \citenamefont {Smet}, \citenamefont {von Klitzing},\ and\ \citenamefont
  {Yacoby}}]{martin-natphys4}%
  \BibitemOpen
  \bibfield  {author} {\bibinfo {author} {\bibfnamefont {J.}~\bibnamefont
  {Martin}}, \bibinfo {author} {\bibfnamefont {N.}~\bibnamefont {Akerman}},
  \bibinfo {author} {\bibfnamefont {G.}~\bibnamefont {Ulbricht}}, \bibinfo
  {author} {\bibfnamefont {T.}~\bibnamefont {Lohmann}}, \bibinfo {author}
  {\bibfnamefont {J.~H.}\ \bibnamefont {Smet}}, \bibinfo {author}
  {\bibfnamefont {K.}~\bibnamefont {von Klitzing}}, \ and\ \bibinfo {author}
  {\bibfnamefont {A.}~\bibnamefont {Yacoby}},\ }\href {\doibase
  10.1038/nphys781} {\bibfield  {journal} {\bibinfo  {journal} {Nat. Phys.}\
  }\textbf {\bibinfo {volume} {4}},\ \bibinfo {pages} {144} (\bibinfo {year}
  {2008})}\BibitemShut {NoStop}%
\bibitem [{\citenamefont {Zhang}\ \emph {et~al.}(2009)\citenamefont {Zhang},
  \citenamefont {Brar}, \citenamefont {Girit}, \citenamefont {Zettl},\ and\
  \citenamefont {Crommie}}]{zhang-natphys5}%
  \BibitemOpen
  \bibfield  {author} {\bibinfo {author} {\bibfnamefont {Y.}~\bibnamefont
  {Zhang}}, \bibinfo {author} {\bibfnamefont {V.~W.}\ \bibnamefont {Brar}},
  \bibinfo {author} {\bibfnamefont {C.}~\bibnamefont {Girit}}, \bibinfo
  {author} {\bibfnamefont {A.}~\bibnamefont {Zettl}}, \ and\ \bibinfo {author}
  {\bibfnamefont {M.~F.}\ \bibnamefont {Crommie}},\ }\href {\doibase
  10.1038/nphys1365} {\bibfield  {journal} {\bibinfo  {journal} {Nat. Phys.}\
  }\textbf {\bibinfo {volume} {5}},\ \bibinfo {pages} {722} (\bibinfo {year}
  {2009})}\BibitemShut {NoStop}%
\bibitem [{\citenamefont {Deshpande}\ \emph {et~al.}(2009)\citenamefont
  {Deshpande}, \citenamefont {Bao}, \citenamefont {Miao}, \citenamefont {Lau},\
  and\ \citenamefont {LeRoy}}]{deshpande-prb79}%
  \BibitemOpen
  \bibfield  {author} {\bibinfo {author} {\bibfnamefont {A.}~\bibnamefont
  {Deshpande}}, \bibinfo {author} {\bibfnamefont {W.}~\bibnamefont {Bao}},
  \bibinfo {author} {\bibfnamefont {F.}~\bibnamefont {Miao}}, \bibinfo {author}
  {\bibfnamefont {C.~N.}\ \bibnamefont {Lau}}, \ and\ \bibinfo {author}
  {\bibfnamefont {B.~J.}\ \bibnamefont {LeRoy}},\ }\href {\doibase
  10.1103/PhysRevB.79.205411} {\bibfield  {journal} {\bibinfo  {journal} {Phys.
  Rev. B}\ }\textbf {\bibinfo {volume} {79}},\ \bibinfo {pages} {205411}
  (\bibinfo {year} {2009})}\BibitemShut {NoStop}%
\bibitem [{\citenamefont {Rossi}\ and\ \citenamefont
  {Das~Sarma}(2008)}]{rossi-prl101}%
  \BibitemOpen
  \bibfield  {author} {\bibinfo {author} {\bibfnamefont {E.}~\bibnamefont
  {Rossi}}\ and\ \bibinfo {author} {\bibfnamefont {S.}~\bibnamefont
  {Das~Sarma}},\ }\href {\doibase 10.1103/PhysRevLett.101.166803} {\bibfield
  {journal} {\bibinfo  {journal} {Phys. Rev. Lett.}\ }\textbf {\bibinfo
  {volume} {101}},\ \bibinfo {pages} {166803} (\bibinfo {year}
  {2008})}\BibitemShut {NoStop}%
\bibitem [{\citenamefont {Polini}\ \emph {et~al.}(2008)\citenamefont {Polini},
  \citenamefont {Tomadin}, \citenamefont {Asgari},\ and\ \citenamefont
  {MacDonald}}]{polini-prb78}%
  \BibitemOpen
  \bibfield  {author} {\bibinfo {author} {\bibfnamefont {M.}~\bibnamefont
  {Polini}}, \bibinfo {author} {\bibfnamefont {A.}~\bibnamefont {Tomadin}},
  \bibinfo {author} {\bibfnamefont {R.}~\bibnamefont {Asgari}}, \ and\ \bibinfo
  {author} {\bibfnamefont {A.~H.}\ \bibnamefont {MacDonald}},\ }\href {\doibase
  10.1103/PhysRevB.78.115426} {\bibfield  {journal} {\bibinfo  {journal} {Phys.
  Rev. B}\ }\textbf {\bibinfo {volume} {78}},\ \bibinfo {pages} {115426}
  (\bibinfo {year} {2008})}\BibitemShut {NoStop}%
\bibitem [{\citenamefont {V\'azquez~de Parga}\ \emph
  {et~al.}(2008)\citenamefont {V\'azquez~de Parga}, \citenamefont {Calleja},
  \citenamefont {Borca}, \citenamefont {Passeggi}, \citenamefont {Hinarejos},
  \citenamefont {Guinea},\ and\ \citenamefont
  {Miranda}}]{vazquezdeparga-prl100}%
  \BibitemOpen
  \bibfield  {author} {\bibinfo {author} {\bibfnamefont {A.~L.}\ \bibnamefont
  {V\'azquez~de Parga}}, \bibinfo {author} {\bibfnamefont {F.}~\bibnamefont
  {Calleja}}, \bibinfo {author} {\bibfnamefont {B.}~\bibnamefont {Borca}},
  \bibinfo {author} {\bibfnamefont {M.~C.~G.}\ \bibnamefont {Passeggi}},
  \bibinfo {author} {\bibfnamefont {J.~J.}\ \bibnamefont {Hinarejos}}, \bibinfo
  {author} {\bibfnamefont {F.}~\bibnamefont {Guinea}}, \ and\ \bibinfo {author}
  {\bibfnamefont {R.}~\bibnamefont {Miranda}},\ }\href {\doibase
  10.1103/PhysRevLett.100.056807} {\bibfield  {journal} {\bibinfo  {journal}
  {Phys. Rev. Lett.}\ }\textbf {\bibinfo {volume} {100}},\ \bibinfo {pages}
  {056807} (\bibinfo {year} {2008})}\BibitemShut {NoStop}%
\bibitem [{\citenamefont {Gibertini}\ \emph {et~al.}(2012)\citenamefont
  {Gibertini}, \citenamefont {Tomadin}, \citenamefont {Guinea}, \citenamefont
  {Katsnelson},\ and\ \citenamefont {Polini}}]{gibertini-prb85}%
  \BibitemOpen
  \bibfield  {author} {\bibinfo {author} {\bibfnamefont {M.}~\bibnamefont
  {Gibertini}}, \bibinfo {author} {\bibfnamefont {A.}~\bibnamefont {Tomadin}},
  \bibinfo {author} {\bibfnamefont {F.}~\bibnamefont {Guinea}}, \bibinfo
  {author} {\bibfnamefont {M.~I.}\ \bibnamefont {Katsnelson}}, \ and\ \bibinfo
  {author} {\bibfnamefont {M.}~\bibnamefont {Polini}},\ }\href {\doibase
  10.1103/PhysRevB.85.201405} {\bibfield  {journal} {\bibinfo  {journal} {Phys.
  Rev. B}\ }\textbf {\bibinfo {volume} {85}},\ \bibinfo {pages} {201405}
  (\bibinfo {year} {2012})}\BibitemShut {NoStop}%
\bibitem [{Note1()}]{Note1}%
  \BibitemOpen
  \bibinfo {note} {The width of the density fluctuations in graphene vary \cite
  {xue-natmat10} from $\sim 10^{10}\protect \mathrm {cm}^{-2}$ (best samples on
  hBN) to $\sim 10^{11}\protect \mathrm {cm}^{-2}$ (on SiO$_2$).}\BibitemShut
  {Stop}%
\bibitem [{\citenamefont {Fulde}\ and\ \citenamefont
  {Ferrell}(1964)}]{fulde-pr135}%
  \BibitemOpen
  \bibfield  {author} {\bibinfo {author} {\bibfnamefont {P.}~\bibnamefont
  {Fulde}}\ and\ \bibinfo {author} {\bibfnamefont {R.~A.}\ \bibnamefont
  {Ferrell}},\ }\href {\doibase 10.1103/PhysRev.135.A550} {\bibfield  {journal}
  {\bibinfo  {journal} {Phys. Rev.}\ }\textbf {\bibinfo {volume} {135}},\
  \bibinfo {pages} {A550} (\bibinfo {year} {1964})}\BibitemShut {NoStop}%
\bibitem [{\citenamefont {Das~Sarma}\ \emph {et~al.}(2007)\citenamefont
  {Das~Sarma}, \citenamefont {Hwang},\ and\ \citenamefont
  {Tse}}]{dassarma-prb75}%
  \BibitemOpen
  \bibfield  {author} {\bibinfo {author} {\bibfnamefont {S.}~\bibnamefont
  {Das~Sarma}}, \bibinfo {author} {\bibfnamefont {E.~H.}\ \bibnamefont
  {Hwang}}, \ and\ \bibinfo {author} {\bibfnamefont {W.-K.}\ \bibnamefont
  {Tse}},\ }\href {\doibase 10.1103/PhysRevB.75.121406} {\bibfield  {journal}
  {\bibinfo  {journal} {Phys. Rev. B}\ }\textbf {\bibinfo {volume} {75}},\
  \bibinfo {pages} {121406} (\bibinfo {year} {2007})}\BibitemShut {NoStop}%
\bibitem [{\citenamefont {{Das Sarma}}\ and\ \citenamefont
  {{Hwang}}(2012)}]{dassarma-arXiv1203}%
  \BibitemOpen
  \bibfield  {author} {\bibinfo {author} {\bibfnamefont {S.}~\bibnamefont {{Das
  Sarma}}}\ and\ \bibinfo {author} {\bibfnamefont {E.~H.}\ \bibnamefont
  {{Hwang}}},\ }\href@noop {} {\bibfield  {journal} {\bibinfo  {journal} {ArXiv
  e-prints}\ } (\bibinfo {year} {2012})},\ \Eprint
  {http://arxiv.org/abs/1203.2627} {arXiv:1203.2627 [cond-mat.mes-hall]}
  \BibitemShut {NoStop}%
\bibitem [{Note2()}]{Note2}%
  \BibitemOpen
  \bibinfo {note} {Throughout this paper, we will use $\kappa =3.9$
  corresponding to the dielectric constant for hBN or SiO$_2$.}\BibitemShut
  {Stop}%
\bibitem [{\citenamefont {Hwang}\ and\ \citenamefont
  {Das~Sarma}(2007)}]{hwang-prb75}%
  \BibitemOpen
  \bibfield  {author} {\bibinfo {author} {\bibfnamefont {E.~H.}\ \bibnamefont
  {Hwang}}\ and\ \bibinfo {author} {\bibfnamefont {S.}~\bibnamefont
  {Das~Sarma}},\ }\href {\doibase 10.1103/PhysRevB.75.205418} {\bibfield
  {journal} {\bibinfo  {journal} {Phys. Rev. B}\ }\textbf {\bibinfo {volume}
  {75}},\ \bibinfo {pages} {205418} (\bibinfo {year} {2007})}\BibitemShut
  {NoStop}%
\bibitem [{\citenamefont {Clogston}(1962)}]{clogston-prl9}%
  \BibitemOpen
  \bibfield  {author} {\bibinfo {author} {\bibfnamefont {A.~M.}\ \bibnamefont
  {Clogston}},\ }\href {\doibase 10.1103/PhysRevLett.9.266} {\bibfield
  {journal} {\bibinfo  {journal} {Phys. Rev. Lett.}\ }\textbf {\bibinfo
  {volume} {9}},\ \bibinfo {pages} {266} (\bibinfo {year} {1962})}\BibitemShut
  {NoStop}%
\bibitem [{\citenamefont {Sheehy}\ and\ \citenamefont
  {Radzihovsky}(2007)}]{sheehy-annphys322}%
  \BibitemOpen
  \bibfield  {author} {\bibinfo {author} {\bibfnamefont {D.~E.}\ \bibnamefont
  {Sheehy}}\ and\ \bibinfo {author} {\bibfnamefont {L.}~\bibnamefont
  {Radzihovsky}},\ }\href@noop {} {\bibfield  {journal} {\bibinfo  {journal}
  {Annuls of Phys.}\ }\textbf {\bibinfo {volume} {322}},\ \bibinfo {pages}
  {1790} (\bibinfo {year} {2007})}\BibitemShut {NoStop}%
\bibitem [{\citenamefont {Conduit}\ \emph {et~al.}(2008)\citenamefont
  {Conduit}, \citenamefont {Conlon},\ and\ \citenamefont
  {Simons}}]{conduit-pra77}%
  \BibitemOpen
  \bibfield  {author} {\bibinfo {author} {\bibfnamefont {G.~J.}\ \bibnamefont
  {Conduit}}, \bibinfo {author} {\bibfnamefont {P.~H.}\ \bibnamefont {Conlon}},
  \ and\ \bibinfo {author} {\bibfnamefont {B.~D.}\ \bibnamefont {Simons}},\
  }\href {\doibase 10.1103/PhysRevA.77.053617} {\bibfield  {journal} {\bibinfo
  {journal} {Phys. Rev. A}\ }\textbf {\bibinfo {volume} {77}},\ \bibinfo
  {pages} {053617} (\bibinfo {year} {2008})}\BibitemShut {NoStop}%
\bibitem [{\citenamefont {Abergel}\ \emph {et~al.}(2011)\citenamefont
  {Abergel}, \citenamefont {Hwang},\ and\ \citenamefont
  {Das~Sarma}}]{abergel-prb83}%
  \BibitemOpen
  \bibfield  {author} {\bibinfo {author} {\bibfnamefont {D.~S.~L.}\
  \bibnamefont {Abergel}}, \bibinfo {author} {\bibfnamefont {E.~H.}\
  \bibnamefont {Hwang}}, \ and\ \bibinfo {author} {\bibfnamefont
  {S.}~\bibnamefont {Das~Sarma}},\ }\href {\doibase 10.1103/PhysRevB.83.085429}
  {\bibfield  {journal} {\bibinfo  {journal} {Phys. Rev. B}\ }\textbf {\bibinfo
  {volume} {83}},\ \bibinfo {pages} {085429} (\bibinfo {year}
  {2011})}\BibitemShut {NoStop}%
\bibitem [{\citenamefont {Xue}\ \emph {et~al.}(2011)\citenamefont {Xue},
  \citenamefont {Sanchez-Yamagishi}, \citenamefont {Bulmash}, \citenamefont
  {Jacquod}, \citenamefont {Deshpande}, \citenamefont {Watanabe}, \citenamefont
  {Taniguchi}, \citenamefont {Jarillo-Herrero},\ and\ \citenamefont
  {LeRoy}}]{xue-natmat10}%
  \BibitemOpen
  \bibfield  {author} {\bibinfo {author} {\bibfnamefont {J.}~\bibnamefont
  {Xue}}, \bibinfo {author} {\bibfnamefont {J.}~\bibnamefont
  {Sanchez-Yamagishi}}, \bibinfo {author} {\bibfnamefont {D.}~\bibnamefont
  {Bulmash}}, \bibinfo {author} {\bibfnamefont {P.}~\bibnamefont {Jacquod}},
  \bibinfo {author} {\bibfnamefont {A.}~\bibnamefont {Deshpande}}, \bibinfo
  {author} {\bibfnamefont {K.}~\bibnamefont {Watanabe}}, \bibinfo {author}
  {\bibfnamefont {T.}~\bibnamefont {Taniguchi}}, \bibinfo {author}
  {\bibfnamefont {P.}~\bibnamefont {Jarillo-Herrero}}, \ and\ \bibinfo {author}
  {\bibfnamefont {B.~J.}\ \bibnamefont {LeRoy}},\ }\href {\doibase
  10.1038/nmat2968} {\bibfield  {journal} {\bibinfo  {journal} {Nat. Mater.}\
  }\textbf {\bibinfo {volume} {10}},\ \bibinfo {pages} {282} (\bibinfo {year}
  {2011})}\BibitemShut {NoStop}%
\bibitem [{\citenamefont {Rossi}\ and\ \citenamefont
  {Das~Sarma}(2011)}]{rossi-prl107}%
  \BibitemOpen
  \bibfield  {author} {\bibinfo {author} {\bibfnamefont {E.}~\bibnamefont
  {Rossi}}\ and\ \bibinfo {author} {\bibfnamefont {S.}~\bibnamefont
  {Das~Sarma}},\ }\href {\doibase 10.1103/PhysRevLett.107.155502} {\bibfield
  {journal} {\bibinfo  {journal} {Phys. Rev. Lett.}\ }\textbf {\bibinfo
  {volume} {107}},\ \bibinfo {pages} {155502} (\bibinfo {year}
  {2011})}\BibitemShut {NoStop}%
\bibitem [{\citenamefont {Dean}\ \emph {et~al.}(2010)\citenamefont {Dean},
  \citenamefont {Young}, \citenamefont {Meric}, \citenamefont {Lee},
  \citenamefont {Wang}, \citenamefont {Sorgenfrei}, \citenamefont {Watanabe},
  \citenamefont {Taniguchi}, \citenamefont {Kim}, \citenamefont {Shepard},\
  and\ \citenamefont {Hone}}]{dean-natnano5}%
  \BibitemOpen
  \bibfield  {author} {\bibinfo {author} {\bibfnamefont {C.~R.}\ \bibnamefont
  {Dean}}, \bibinfo {author} {\bibfnamefont {A.~F.}\ \bibnamefont {Young}},
  \bibinfo {author} {\bibfnamefont {I.}~\bibnamefont {Meric}}, \bibinfo
  {author} {\bibfnamefont {C.}~\bibnamefont {Lee}}, \bibinfo {author}
  {\bibfnamefont {L.}~\bibnamefont {Wang}}, \bibinfo {author} {\bibfnamefont
  {S.}~\bibnamefont {Sorgenfrei}}, \bibinfo {author} {\bibfnamefont
  {K.}~\bibnamefont {Watanabe}}, \bibinfo {author} {\bibfnamefont
  {T.}~\bibnamefont {Taniguchi}}, \bibinfo {author} {\bibfnamefont
  {P.}~\bibnamefont {Kim}}, \bibinfo {author} {\bibfnamefont {K.~L.}\
  \bibnamefont {Shepard}}, \ and\ \bibinfo {author} {\bibfnamefont
  {J.}~\bibnamefont {Hone}},\ }\href {\doibase 10.1038/nnano.2010.172}
  {\bibfield  {journal} {\bibinfo  {journal} {Nat Nano}\ }\textbf {\bibinfo
  {volume} {5}},\ \bibinfo {pages} {722} (\bibinfo {year} {2010})}\BibitemShut
  {NoStop}%
\bibitem [{Note3()}]{Note3}%
  \BibitemOpen
  \bibinfo {note} {Such experiments could benefit enormously if the
  inhomogeneity in the individual graphene layers could be screened out in a
  controlled manner by using screening layers as was done in a clever recent
  experiment by Ponomarenko {\protect \textit {et al.}} .\cite
  {ponomarenko-natphys7}}\BibitemShut {NoStop}%
\bibitem [{\citenamefont {Ponomarenko}\ \emph {et~al.}(2011)\citenamefont
  {Ponomarenko}, \citenamefont {Geim}, \citenamefont {Zhukov}, \citenamefont
  {Jalil}, \citenamefont {Morozov}, \citenamefont {Novoselov}, \citenamefont
  {Grigorieva}, \citenamefont {Hill}, \citenamefont {Cheianov}, \citenamefont
  {Fal'ko}, \citenamefont {Watanabe}, \citenamefont {Taniguchi},\ and\
  \citenamefont {Gorbachev}}]{ponomarenko-natphys7}%
  \BibitemOpen
  \bibfield  {author} {\bibinfo {author} {\bibfnamefont {L.~A.}\ \bibnamefont
  {Ponomarenko}}, \bibinfo {author} {\bibfnamefont {A.~K.}\ \bibnamefont
  {Geim}}, \bibinfo {author} {\bibfnamefont {A.~A.}\ \bibnamefont {Zhukov}},
  \bibinfo {author} {\bibfnamefont {R.}~\bibnamefont {Jalil}}, \bibinfo
  {author} {\bibfnamefont {S.~V.}\ \bibnamefont {Morozov}}, \bibinfo {author}
  {\bibfnamefont {K.~S.}\ \bibnamefont {Novoselov}}, \bibinfo {author}
  {\bibfnamefont {I.~V.}\ \bibnamefont {Grigorieva}}, \bibinfo {author}
  {\bibfnamefont {E.~H.}\ \bibnamefont {Hill}}, \bibinfo {author}
  {\bibfnamefont {V.~V.}\ \bibnamefont {Cheianov}}, \bibinfo {author}
  {\bibfnamefont {V.~I.}\ \bibnamefont {Fal'ko}}, \bibinfo {author}
  {\bibfnamefont {K.}~\bibnamefont {Watanabe}}, \bibinfo {author}
  {\bibfnamefont {T.}~\bibnamefont {Taniguchi}}, \ and\ \bibinfo {author}
  {\bibfnamefont {R.~V.}\ \bibnamefont {Gorbachev}},\ }\href {\doibase
  10.1038/nphys2114} {\bibfield  {journal} {\bibinfo  {journal} {Nat Phys}\
  }\textbf {\bibinfo {volume} {7}},\ \bibinfo {pages} {958} (\bibinfo {year}
  {2011})}\BibitemShut {NoStop}%
\end{thebibliography}%

\end{document}